\documentstyle[12pt]{article}
\advance\textheight by \topskip
\begin{document}
\begin{flushright}
SU-ITP-93-16\\
hepth@xxx/uummnnn
\end{flushright}
\vspace{1.7 cm}
\begin{center}
{\LARGE\bf Killing Spinor Identities}\\
\vskip 1.8 cm
\bf Renata Kallosh \footnote {E-mail address:
kallosh@physics.stanford.edu}
 and
Tom\'as Ort\'\i n \footnote{E-mail address: tomaso@slac.stanford.edu}
 \vskip 0.05cm
{\it Physics Department, Stanford University, Stanford,   CA 94305}
\end{center}
\vskip .5 cm
\centerline{\bf ABSTRACT}
\begin{quotation}

We have found generic Killing spinor identities which bosonic
equations of
motion have to satisfy in supersymmetric theories if the solutions
admit Killing
spinors. Those identities constrain possible quantum corrections to
bosonic solutions with unbroken supersymmetries.

As an application we show that purely electric static extreme dilaton
black holes may acquire specific quantum corrections, but  the purely
magnetic ones cannot.
\end{quotation}

\normalsize
\newpage
In recent years there was a strong interest in finding exact
classical solutions in theories including gravity.
Some of these solutions have very interesting properties, which
suggest that they remain exact even after  quantum
corrections are taken into account.
Non-renormalization theorem  were used to show that some solutions
do not acquire  quantum corrections.
In this article we will show that in addition to the methods used
before there exists a new method, which can be used in order to
find and investigate classical solutions, which are not affected by
quantum corrections. In some special cases, where the solutions
possess Killing spinors, there exist identities, which serve as a
powerful tool  which help to control quantum
corrections.

In order to explain the origin of the new identities, let us consider
a classical Lagrange equation derived from  some action principle.
\begin{equation}
{\delta S \over \delta \phi ^b}\equiv   S,_b =  0\ .
\end{equation}
Quantum corrections modify those equations by introducing
a right-hand side,
which is absent  in the classical approximation:
\begin{equation}\label{quant}
   S,_b =  J_b\ .
\end{equation}
Generically, no information is available about the quantum
corrections to
the equations
of motion in 4-dimensional gravitational problems. We will exhibit
 here some particular
 situations when such information is available due to symmetry
properties of
$J_b$, which follow from the symmetry properties of $S,_b$.
A well known example is given by the Einstein equations
\begin{equation}
R_{\mu\nu} - {1\over2} g_{\mu\nu} R = T_{\mu\nu}\ .
\label{E}\end{equation}
The stress-energy tensor
$T_{\mu\nu}$ appears in the
right-hand side of Einstein equations
and  may describe both classical sources as well as quantum
corrections
to classical equations. The left-hand side of Einstein equations
satisfies the identities
\begin{equation}
\nabla^{\mu} \; (R_{\mu\nu} - {1\over2} g_{\mu\nu} R) = 0\ .
\label{iden}\end{equation}
The
right-hand side of Einstein equations also must satisfy the covariant
conservation law
\begin{equation}
\nabla^{\mu} \; T_{\mu\nu} = 0 \ .
\label{cons}\end{equation}
This law was used in numerous problems of general relativity. Even
when
$T_{\mu\nu}$ is not known, some of its properties can be deduced from
its covariant conservation law, supplemented by specific assumptions
like time independence, etc.

Historically it was the other way around. In his first attempt
\cite{EIN1} to find
the field equations for gravity Einstein proposed
\begin{equation}
R_{\mu\nu}  = T_{\mu\nu}\ .
\end{equation}
He realized very soon  that
 in general this equation contradicts the covariant conservation of
the energy
momentum
tensor. The first paper was
submitted
on November 11. On November 18 Einstein submitted an {\it Addendum}
\cite{EIN2} where he
required that  $T_\mu{}^\mu = 0$ for
consistency with conservation of energy and momentum. Finally, on
December 2 he submitted the second paper \cite{EIN3} where the
crucial final step  was made to introduce \footnote{Actually he first
introduced it in the form $R_{\mu\nu} = T_{\mu\nu}  - {1\over2}
g_{\mu\nu}
T$.} eq. (\ref{E}). It was also realized  that when Einstein
equations are derived
from
 the Einstein-Hilbert   Lagrangian, the left hand side of eq.
(\ref{E})
appears automatically, and the identity (\ref{iden}) is the
consequence of the
general covariance of the  action. Thus, either one knows some
identities which the right-hand side of the equations satisfies
 due to some symmetry
 principle
(Einstein started with
$\nabla^{\mu} \; T_{\mu\nu} = 0 )
$ and requires the same identities to be satisfied by the
left-hand-side
(which enforces general covariance in Einstein's theory)
or vice versa.

The purpose of this paper is to show that for bosonic
configurations of
supersymmetric theories with unbroken supersymmetries,
the equations of motion satisfy some generic identities. These
identities
 are due to the symmetries of the theory and the solutions.
 This forces the right-hand side of these equations,
coming
both from classical sources and from quantum corrections, to satisfy
the same identities. An
investigation of these identities for specific configurations may
give a
substantial information about the properties of quantum corrections
 which may differ from one configuration to another.

 We will first rewrite the information discussed above in a
language which will
be easy generalizable to our problem.

The fact that the action $S = \int d^4 x\sqrt {-g} \;R $ is general
covariant means that

\begin{equation}
\delta_{\xi} S\equiv  {\delta S \over \delta g^{\mu\nu}}
\delta_{\xi}\;
g^{\mu\nu}= 0 \ ,
\label{inv}
\end{equation}
where $\delta_{\xi}\; g^{\mu\nu}\equiv 2\nabla^{(\mu}
\xi^{\nu)}$ is the general covariance transformation of the metric.
Let us show that eq. (\ref{inv}) does imply the identity
(\ref{iden}). Indeed, in
detailed form eq. (\ref{inv}) is
\begin{equation}
\int d^4 x\; {\delta S \over \delta g^{\mu\nu}} \nabla^{\mu}
\xi^{\nu}= \int d^4 x\sqrt {-g}\;( R_{\mu\nu} - {1\over2} g_{\mu\nu}
R)
\nabla^{\mu} \xi^{\nu} = 0 \ .
\label{iden2}\end{equation}
Since $\xi^{\nu}$ is an arbitrary function of $x$, we may vary
equation
(\ref{inv}) over $\xi^{\nu}$. The result is
\begin{equation}
{\delta S \over \delta \xi^{\nu} (x)}  =
-2\sqrt {-g} \; \nabla^{\mu}
 ( R_{\mu\nu} - {1\over2} g_{\mu\nu} R) = 0 \ ,
\label{var}\end{equation}
which is equivalent to eq. (\ref{iden}). Thus it is the symmetry of
the action,
which is responsible for the fact that some mix of the variational
derivatives of
the action over the metric, as given in eq. (\ref{inv}), has to be
zero. Therefore
if one wants to consider the equation
\begin{equation}
{\delta S \over \delta g^{\mu\nu}} =  \sqrt {-g}\; T_{\mu\nu}\ ,
\end{equation}
the right-hand side of this equation also has to satisfy the
identity, following
from replacing ${\delta S \over \delta g^{\mu\nu}}$ in eq.
(\ref{iden2}) by
$\sqrt {-g}\; T_{\mu\nu}$. The covariant conservation of the
energy-momentum
tensor (\ref{cons}) follows.

Let us start with some action $S$ which has local supersymmetry.
This means that there
exist some supersymmetric transformations $\delta_{\epsilon} \,
\phi^b$,
 $\delta_{\epsilon}\,\phi^f$
with parameter $\epsilon(x)$ of all bosonic fields $\phi^b$ and of
all
fermionic fields $\phi^f$
such that
\begin{equation}
\delta_{\epsilon} S  \equiv
 \sum_{b}  S,_b \; \delta_{\epsilon}\,\phi^b
  \hskip 0.5 cm +  \sum_{f} S,_f \; \delta_{\epsilon}\,\phi^f = 0\ .
\label{supersym}\end{equation}
This identity relates all variational derivatives of the action over
bosons
($  {\delta S \over \delta \phi^b} \equiv S,_{b}$) to those over
fermions
($  {\delta S \over \delta \phi^f} \equiv S,_f$).

Identity (\ref{supersym}) is satisfied for any (on- or off-shell)
values of bosonic and
fermionic
fields and arbitrary local fermionic parameter
$\epsilon(x)$. We would like to vary this identity
over the fermions once, and after  the variation to
set all fermions to zero. It is useful to remember that $ S,_b$ and $
\delta_{\epsilon} \, \phi^f$ are even functions of fermions, and $
S,_f$ and $
\delta_{\epsilon} \, \phi^b $ are odd functions of fermions.
Therefore we get
\begin{equation}
(\delta_{\epsilon} S),_{f_2}  \equiv
\left.\left( \sum_{b}  S,_b \;(\delta_{\epsilon}\,\phi^b) ,_{f_2}
  \hskip 0.5 cm +  \sum_{f_1} S,_{f_1f_2}
\;\delta_{\epsilon}\,\phi^f \right) \right|_{\phi^f = 0} = 0\ .
\label{bos1}\end{equation}
In this form this identity is still not very useful for the purely
bosonic
part
of the system,
in which we are interested. One may expect that quantum corrections
will change the
bosonic equations from $S,_b = 0$ to
$S,_b = J_b\ .$
We want to extract some information about the
left-hand side
of this equation from symmetries of the theory and use them as the
constraints
on the possible form of $J_b$.

 Equation (\ref{bos1}), which follows only from the symmetries of the
theory,
is purely bosonic, but still  in addition
to bosonic variational derivatives $S,_b$ of the action the second
variational
derivatives of the action over fermions $S,_{f_1f_2} \equiv {\delta^2
S \over \delta \phi^{f_1}\delta \phi^{f_2}}$ enters in
the second term. Therefore we have to impose some additional
assumption about
the configuration for which we are going to study the proposed
equation
(\ref{quant}): if we consider only bosonic configurations admitting
 Killing spinors (KS), i. e. supersymmetry parameters $\epsilon(x)$
 satisfying
\begin{equation}
\left. \delta_{\epsilon_{\phantom{}_{{Killing}}}} \, \phi^f
\right|_{\phi^f = 0} = 0\ ,
\label{killing} \end{equation}
then we are left with the following
identities
\begin{equation}
(\delta_{\epsilon_{\phantom{}_{{Killing}}}} S) ,_{f}  \equiv
\left. \sum_{b}  S,_b \;(\delta_{\epsilon_{\phantom{}_{{Killing}}}}
\, \phi^b),_{f }
\right|_{\phi^f = 0} = 0\ .
\label{bos2}
\end{equation}
Our proposed equation (\ref{quant}) with quantum corrections taken
into
account is consistent only if

\begin{equation}
\left. \sum_{b}  S,_b \;(\delta_{\epsilon_{\phantom{}_{{Killing}}}}
\, \phi^b) ,_{f}
  \right|_{\phi^f = 0} =
\left. \sum_{b}  J_b \;(\delta_{\epsilon_{\phantom{}_{{Killing}}}} \,
\phi^b) ,_{f}
 \right|_{\phi^f = 0} = 0\ .
\label{KI}
\end{equation}

We will call these identities {\it Killing Spinor Identities} (KSI).
Although they were
derived from supersymmetry, they are identities for bosonic fields
only.
In words, equations (\ref{KI}) tell us that one has to check the
consistency of eqs.
(\ref{quant}) as follows: For every bosonic field of the theory one
has to
multiply the  $J_b$-term by the corresponding bosonic
supersymmetry transformation with the KS
as a parameter, and take a sum over all
bosons. The sum, varied over each fermionic field, should vanish!

Equation (\ref{KI}) is our main result. It is interesting that in
numerous
studies of bosonic solutions of equations of motions which have
unbroken
supersymmetries, i.e. admit KS's, only fermionic
transformations
were used to establish the configuration for which eq.
(\ref{killing}) is
satisfied. Our KSI may help to find the most
general
solution of KS equations (\ref{killing}), which either permits
the
non-trivial $J_b$ terms in the right-hand side of eq. (\ref{quant})
or forbids
them. In the first case quantum corrections to supersymmetric bosonic
solutions of classical field equations are possible, in the second
case there
are no such corrections.  {\it  The additional input comes from the
knowledge of the
supersymmetric transformation rules for bosonic fields, which was not
used
before in the context of bosonic configurations with unbroken
supersymmetries}.

We would like to apply the KSI (\ref{KI}) to the
configuration of  $U(1)$ dilaton and axion-dilaton 4d black holes. We
do
not want to use the already known  extreme solutions of classical
field equations
\cite{G}, \cite{US}. Those solutions  have some unbroken
supersymmetries \cite{US},
but our purpose here is to find whether the KS equations
(\ref{killing})
have more general solutions than those which solve exactly the
classical field
equations. The action with $N=4$ local supersymmetry is the $SU(4)$
version
of $N=4$ supergravity \cite{CSF}. Its bosonic part, which will
be considered here, depends on the vierbein
$g_{\mu\nu}$, on the abelian vector field\footnote{We will consider
only solutions with one vector field
for simplicity and take, as in \cite{US}, $A_{\mu}= A_{\mu}{}^3$.}
$A_{\mu}$
and on the dilaton-axion field
$\lambda = a + i e^{-2\phi}$ .  Our notation
is defined in \cite{US}.
The equations of motion, which may acquire
a non-trivial right-hand side due to quantum corrections are
\begin{eqnarray}
{\delta S^{cl} \over \delta g_{\mu\nu}}&=& J^{\mu\nu}\ , \hskip 2 cm
{\delta S^{cl} \over \delta \lambda}= J\ ,\nonumber\\
\nonumber\\
{\delta S^{cl} \over \delta A_{\mu}} &=& J^{\mu}\ , \hskip 2,1 cm
{\delta S^{cl} \over \delta \bar \lambda}= \bar J \ .
\label{qu.cor}\end{eqnarray}
The axion-dilaton extreme black hole solutions \cite{G}, \cite{US}
are solutions of equations (\ref{qu.cor}) with
\begin{equation}
 J^{\mu\nu}= J^{\mu}= J =  \bar J = 0\ .
\end{equation}
The extreme solutions are supersymmetric \cite{US}, i.e. they admit
KS's (\ref{killing}).
The four gravitinos $\Psi_{\mu I}$ and four dilatinos $\Lambda_{I}$
have
vanishing local supersymmetry transformations in presence of gravity,
vector field  and dilaton.
\begin{eqnarray}
\delta_{\epsilon_{\phantom{}_{{Killing}}}}\Psi_{\mu I}  = 0 \ ,
\hskip 1 cm
\delta_{\epsilon_{\phantom{}_{{Killing}}}}\Lambda_I  = 0\ ,
  \,  \hskip 1 cm  I = 1, 2, 3, 4.
\label{susy}\end{eqnarray}

{\it A priori} it is not known whether there exist solutions of
Killing
equations (\ref{susy}) which simultaneously solve the dynamical
equations
(\ref{qu.cor}) with non-trivial right-hand sides.  Using the KSI
(\ref{KI}) derived above one can address this problem.
One has to take into account the following supersymmetry
transformations
of our bosonic fields\footnote{In this particular example we do not
consider
$\alpha^{\prime}$ corrections to the $N=4$ supersymmetry
transformation rules. However, the general Killing Spinor Identities
given in
eq. (\ref{KI}) are correct also when one includes them.}
\begin{eqnarray}
\delta_{\epsilon}\, g_{\mu\nu} & =& \bar \epsilon^I   \gamma_{\mu}
\Psi_{\nu I
} +
\bar \epsilon_I   \gamma_{\mu }  \Psi_{\nu}{}^I +
\bar \epsilon^I   \gamma_\nu   \Psi_{\mu I
} +
\bar \epsilon_I   \gamma_\nu   \Psi_{\mu}{}^I  \ ,\nonumber\\
\nonumber\\
\delta_{\epsilon} \, A_{\mu} & =& - {1\over \sqrt{2}} e^{\phi}
(\bar \epsilon^I
\alpha_{IJ}   \Psi_{\mu}^J + \bar \epsilon_I  \alpha^{IJ}
\Psi_{\mu J }
-  \bar \epsilon^I   \alpha_{IJ}  \gamma_\mu \Lambda^J
- \bar \epsilon_I   \alpha^{IJ}  \gamma_\mu \Lambda_J ) \ ,
\nonumber\\
\nonumber\\
\delta_{\epsilon} \, \lambda &=& - 4i  e^{-2 \phi} \bar \epsilon^I
\Lambda_I \ ,
\hskip 2 cm  \delta_{\epsilon} \, \bar \lambda =  4i  e^{-2 \phi}
 \bar \epsilon_I \Lambda^I \ .
\label{transf }\end{eqnarray}
The next step is to form a product of each ``source-term" $J_b$ from
the
right-hand side of eq. (\ref{qu.cor}) with the proper
$\delta_{\epsilon}\, \phi^b$,
i.e. to write the function
\begin{equation}
\Omega \equiv \sum_b  J_b \delta_{\epsilon}\,  \phi^b = J^{\mu\nu}
\delta_{\epsilon}\,
g_{\mu\nu}
+ J^{\mu} \delta_{\epsilon}\, A_{\mu} + {1\over 2}(  J\,
\delta_{\epsilon}\,
 \lambda + c.c.)\ .
\end{equation}
This function is linear in fermions and we have to differentiate it
over all
types of fermions and the result has to vanish. The KSI take the form
$\Omega,_f = 0 \ .$
Let us list these identities for our example.
\begin{eqnarray}
\Omega,_{\Lambda_I}&=& - J^{\mu}  {1\over \sqrt{2}} e^{\phi}
 \bar \epsilon_J   \alpha^{JI}  \gamma_\mu  + 2i J \, e^{-2 \phi}\bar
\epsilon^I  = 0 \ ,\nonumber\\
\nonumber\\
\Omega,_{\Lambda^I} &=& - J^{\mu}  {1\over \sqrt{2}} e^{\phi}
 \bar \epsilon^J   \alpha_{JI}  \gamma_\mu  -  2i \bar J \,
e^{-2\phi}\bar
\epsilon_I  = 0 \ ,\nonumber\\
\nonumber\\
\Omega,_{\Psi_{\mu I}}&=&- 2 J^{\mu\nu} \bar \epsilon^I   \gamma_\nu
+
J^{\mu}  {1\over \sqrt{2}} e^{\phi} \bar \epsilon_J  \alpha^{JI} =
0 \ ,\nonumber\\
\nonumber\\
\Omega,_{\Psi_{\mu}^ I}&=&- 2 J^{\mu\nu} \bar \epsilon_I   \gamma_\nu
+
J^{\mu}  {1\over \sqrt{2}} e^{\phi} \bar \epsilon^J  \alpha_{JI} =
0 \ .
\label{example}\end{eqnarray}
We assume that the KS equations (\ref{susy}) for  our metric, vector
field and axion-dilaton are
are satisfied (i.e. $\epsilon=\epsilon_{Killing}$)
 and that simultaneously these configurations solve the equations
of motion
(\ref{qu.cor}).
 The first pair of equations (\ref{example}) has solutions only if
\begin{equation}
J_{\mu}J^{\mu}  = 8 \,  e^{-6 \phi}|J|^2 \ .
\label{condition1}
\end{equation}

The second pair of equations implies

\begin{equation}
J^{\mu}J^{\nu}  = 8 \,  e^{-2 \phi}J^\mu{}_{\eta}  J^{\eta \nu}\ .
\label{condition2}
\end{equation}

These conditions can be more or less restrictive for different
configurations.
We will study the simplest supersymmetric black hole configurations:
$U(1)$ {\it static dilaton black holes}.
Without axion field ($a=0$, $J= - \bar J = {i\over 2} e^{2 \phi}
J_{\phi}$)
the static black hole solutions are either
electric
($ F_{ij} =0$ and $ J_{i} = 0$) or magnetic ($F_{0i} =0$ and  $J_0 =
0$).
In the first case condition
(\ref{condition1}) is satisfied and
\begin{equation}
  J_{0}^2 = 2 e^{-2 \phi}J_{\phi}^2 \ .
\end{equation}
In the magnetic case equation (\ref{condition1}) cannot be satisfied,
since
the
right-hand side is positive, and the left-hand side is negative.
This proves that the only purely magnetic dilaton black hole with
unbroken
supersymmetries is the well known extreme solution of classical
equations \cite{G}, \cite{US}.
For electric case the solution exists
 under the condition that $J_{a b} $ has only time
components
$J_{0  0} $, and that there exists the relation
\begin{equation}
 |J_{\phi} |  = | J_{0}|  {1\over  \sqrt 2} e^{\phi} = 2 | J_{0  0}|
\ .
\label{kil.con}
\end{equation}
After having established that our KSI are
satisfied
in the electric case with non-vanishing corrections to the classical
equations and in the
magnetic case  with vanishing corrections, we can confirm
this by
looking on the KS equations (\ref{susy}).
The analysis proceeds as the one performed in \cite{US}, however
this time we solve in addition to (\ref{susy}) the  dynamical
equations with some
unknown quantum corrections $J^{\mu}, J_{\phi},  J_{\mu\nu}$:
\begin{eqnarray}
4 \sqrt {-g} \,\nabla_{\mu}(e^{-2\phi}F^{\mu\nu})&=&J^{\mu}
\ ,\nonumber\\
\nonumber\\
- 4 \sqrt {-g} \, (\nabla^{2}\phi -
{\textstyle\frac{1}{2}} e^{-2\phi}F^{2} )&=&J_\phi \ ,
\nonumber\\
\nonumber\\
 -\sqrt {-g} \, \Bigl\lbrace \, R_{\mu\nu} + {\textstyle\frac{1}{2}}
g_{\mu\nu} R
+2 [\nabla_{\mu} \phi \cdot\nabla_{\nu} \phi- {\textstyle\frac{1}{2}}
g_{\mu\nu}  (\nabla \phi)^2]  - &&\nonumber\\
\nonumber\\
-2e^{-2\phi}  [F_{\mu\lambda} F_{\nu\delta} g^{\lambda\delta}
-{\textstyle\frac{1}{4}}
g_{\mu\nu}F^{2}]\Bigr\rbrace
 & = &J_{\mu\nu} \ .\label{mot1}  \end{eqnarray}

The static electric solution of eqs. (\ref{mot1}) and (\ref{susy})
with unbroken $N=2$ supersymmetry exists and is given by
\begin{eqnarray}
ds^{2} &=&  e^{2U}dt^{2}-e^{-2U}d\vec{x}^{2} \ ,\nonumber\\
\nonumber\\
A = \psi dt  &, &F =  d \psi \wedge dt \ , \nonumber\\
\nonumber\\
\sqrt{2}\, \psi = \pm e^{+2U} e^{\phi_0}  &, &\qquad
\phi=U+\phi_{0} \ ,\nonumber\\
\nonumber\\
\partial_i\partial_i \;e^{-2U} &=& -\frac{1}{2}e^{-2U} J_{\phi} \ ,
\label{extsol}
\end{eqnarray}
where the right-hand side of the last equation  is an
arbitrary
function $J_\phi (\vec x)$, which can come either from
external classical sources or from quantum corrections.
For this solution the terms in the right-hand side of equation
(\ref{mot1})
defining corrections to
classical equations are
subject to the constraint $J_\phi = \pm   J_{0}
 {1\over \sqrt{2}} e^{\phi} =  2 J_{0 0}$,
in agreement with the requirement from
the KSI (\ref{kil.con}).

We have checked that the most general static solution of the purely
magnetic type,
which admits KS's of $N=4$ supersymmetry (\ref{susy}) and
solves the
dynamical equations (\ref{mot1}), exists only if
$J^{\mu}= J_{\phi} =  J_{\mu\nu}= 0$, i.e. the extreme magnetic
dilaton black hole cannot acquire quantum corrections, in a complete
agreement
with the prediction from the KSI (\ref{KI}).
Equations
(\ref{susy}) ensure the existence of a dual magnetic potential
\cite{US}, and the solution
is
  \begin{eqnarray}
ds^{2} &=&  e^{2U}dt^{2}-e^{-2U}d\vec{x}^{2} \ ,\nonumber\\
\nonumber\\
\tilde A = \tilde \psi dt  &, &\tilde F = i d \tilde \psi \wedge dt \
, \nonumber\\
\nonumber\\
\sqrt{2}\, \tilde \psi = \pm e^{+2U-\phi_{0}}   &, &\qquad
\phi=-U+\phi_{0} \ ,\nonumber\\
\nonumber\\
\partial_i\partial_i \;e^{-2U} &=& 0 \ .
\label{magn}
\end{eqnarray}

Thus we have found a  rather unexpected difference in the properties
of electric
and magnetic supersymmetric dilaton black hole solutions at the
quantum level. This difference follows from eq.  (\ref{condition1}),
which requires
$J_\mu J^\mu \geq 0$.
In absence of $J^0$ term in the right-hand side of the equation,
defining the
metric, these two solutions are related by $SL(2,R)$-symmetry
\cite{CSF}. But in presence of quantum
corrections the magnetic solution remains the same, the metric
$g^{\hat
0\hat 0} = e^{-2U}$ is still a harmonic function and for the electric
solution
a new unknown function may appear in the right-hand side, and both
types of solutions
have unbroken supersymmetries.\vskip 0.8 cm

Our first experience of using  KSI for the
purpose of investigations
of the non-perturbative  solutions in quantum gravity looks
promising.  More
general black hole solutions, like the ones with axion field may
be tested for quantum corrections.  Any other bosonic configurations
with unbroken
supersymmetries may be tested in the same way.
The KSI presented in eq.
(\ref{KI}) of this paper are valid for any locally supersymmetric
theory in any dimension and with any number of supersymmetries with
or without $\alpha^{\prime}$ corrections.
Thus, each
time equations  $\delta_{\epsilon} \Psi = 0 $ have been used to
find a bosonic
supersymmetric configuration, an additional new information about how
those
configurations are affected by quantum corrections may be available
from Killing spinor identities   (\ref{KI}), applied to each of
those configurations. They involve the {\it specific use of
supersymmetry
transformation rules of bosonic fields of the theory}. As we have
seen, in the black hole case the result is the following: static
purely magnetic
supersymmetric extreme dilaton black hole is not affected by quantum
corrections,
however, the electric one may be affected.  In particular,
we have found specific relations between quantum corrections to
different equations.
We believe that the new method of
using Killing spinor identities in the context of supersymmetric
bosonic configurations improves essentially the quality
and the power of the arguments,
which are used in proving the so-called ``supersymmetric
non-renormalization theorems".

Discussions  with E. Bergshoeff, A. Linde and
A. Van Proeyen  are gratefully acknowledged. The work of R.K. has
been supported by the NSF grant PHY-8612280 and in part by
Stanford University. The work of T.O. has been supported
by a Spanish Government MEC postdoctoral grant.

\end{document}